\documentclass[conference]{IEEEtran}
\ifCLASSINFOpdf
\else
\fi
\usepackage{booktabs}

\usepackage{amsthm,amssymb,graphicx,multirow,amsmath,bm, color,amsfonts}
\usepackage[update,prepend]{epstopdf}
\usepackage{cite}
\usepackage{tabulary}
 
\usepackage{bbm} 

\usepackage{multirow}
\usepackage{comment}
\usepackage[ruled,linesnumbered]{algorithm2e}
\SetKw{KwBy}{by}

\SetKwInput{KwRequire}{Require}
\SetKwInput{KwInit}{Initialize}

\usepackage{siunitx}

\usepackage{makecell}

\DeclareMathOperator*{\argmax}{argmax} 
\DeclareMathOperator*{\argmin}{argmin} 

\def\figref#1{Fig.\,\ref{#1}}%
 
\def\BibTeX{{\rm B\kern-.05em{\sc i\kern-.025em b}\kern-.08em
		T\kern-.1667em\lower.7ex\hbox{E}\kern-.125emX}}
\begin{document}
\pagenumbering{gobble}
\graphicspath{{./figures/}}
\title{
	Channel Estimation in MIMO Systems Using Flow Matching Models
}

\author{
		\IEEEauthorblockN{Yongqiang Zhang\IEEEauthorrefmark{1} and Qurrat-Ul-Ain Nadeem\IEEEauthorrefmark{1}\IEEEauthorrefmark{2}}
		\IEEEauthorblockA{\IEEEauthorrefmark{1}Engineering Division, New York University (NYU) Abu Dhabi,UAE}
		\IEEEauthorblockA{\IEEEauthorrefmark{2}NYU Tandon School of Engineering, New York, USA}
					\IEEEauthorblockA{Email: \{yz12825, qurrat.nadeem\}@nyu.edu}
	}

\maketitle

\begin{abstract}
	Multiple-input multiple-output (MIMO) systems require efficient and accurate channel estimation with low pilot overhead to unlock their full potential for high spectral and energy efficiency.
	While deep generative models have emerged as a powerful foundation for the channel estimation task, the existing approaches using diffusion-based and score-based models suffer from high computational runtime due to their stochastic and many-step iterative sampling.
	In this paper, we introduce a flow matching-based channel estimator to overcome this limitation. 
	The proposed channel estimator is based on a deep neural network trained to learn the velocity field of wireless channels, which we then integrate into a plug-and-play proximal gradient descent (PnP-PGD) framework. 
	Simulation results reveal that our formulated approach consistently outperforms existing state-of-the-art (SOTA) generative model-based estimators, achieves up to 49 times faster inference at test time, and reduces up to 20 times peak graphics processing unit (GPU) memory usage. 
	Our code and models are publicly available to support reproducible research.
\end{abstract}

\section{Introduction}

Estimating accurate channel state information is essential for multiple-input multiple-output (MIMO) communication systems since it serves the backbone for reliable beamforming and efficient resource allocation.
The problem of channel estimation is reconstructing the unknown channel from a limited number of noisy pilot observations, which are corrupted by fading and thermal noise \cite{Marzetta2010}. 
The impracticality of acquiring accurate statistical priors for conventional channel estimators, together with their poor scalability to large arrays, has motivated the exploration of deep learning approaches \textcolor{black}{that formulate} channel estimation as a data-driven supervised regression problem \cite{Soltani2019}.
However, supervised-regression based approaches learn a direct mapping without explicitly capturing the channel's underlying prior distribution, limiting their robustness in low signal-to-noise ratio (SNR) regimes and generalization to unseen channel distributions\cite{Ma2024}.

To address this critical gap in learning the prior distribution, recent progress in unsupervised deep generative modeling has motivated its application in solving the channel estimation problem, which has achieved state-of-the-art (SOTA) performance \cite{Ma2024, Arvinte2023, Zhou2025, Chen2025}.
These approaches can be classified into two categories: score-based models \cite{Arvinte2023, Chen2025} and diffusion-based models \cite{Ma2024, Zhou2025}.
Both categories share the same pipeline: they first define a forward noising process that gradually injects noise into the channel data, and then train a neural network to learn the reverse sampling process that iteratively denoises a noisy measurement to match the prior distribution of the channel dataset.
Specifically, the score-based approaches operate by explicitly learning the \emph{score},  which is defined as the gradient of the log-prior distribution \cite{Song2019}. 
The diffusion-based approaches are trained to reverse this process by directly predicting the noise component that was added at each discrete step \cite{Ho2020}.
However, the sophisticated stochastic nature of the reverse paths learned by these models necessitates a large number of inference steps (typically more than $1000$), thereby incurring high inference latency that hinders their deployment in real-world wireless systems.

By learning a deterministic velocity field to transform a simple distribution into the target data distribution, flow matching models have recently emerged as a promising new paradigm that enables high-fidelity inference in drastically fewer inference steps \cite{Lipman2023,Albergo2022, Liu2022c}.  
This ability to model the data distribution makes flow matching an extremely powerful candidate for a \emph{regularizer} within an optimization framework.
The well-established plug-and-play (PnP) approach is designed to leverage such a \emph{regularizer} by integrating it into a classical iterative algorithm, such as proximal gradient descent (PGD), to solve complex estimation tasks in other domains, such as image restoration \cite{Venkatakrishnan2013, Hurault2022, Ryu2019}.
Motivated by the potential of this synergy, in this paper, we present a learning-based MIMO channel estimator that uses a flow matching model to learn the velocity field of complex MIMO channels and integrates it into a PnP-PGD framework.
To the best of our knowledge, this work is the first attempt to employ flow matching models to solve a problem in wireless communications.
Our main contributions  are summarized as follows:
\begin{itemize} 
	\item 
	A flow matching-based generative model is trained to capture the velocity field governing the intrinsic structure of point-to-point MIMO channels. 
	The training is fully unsupervised and independent of pilot patterns or SNR information, enabling the learned velocity field to generalize across diverse channel conditions and system configurations without retraining.
	\item 
	We develop a PnP-PGD channel estimation framework that integrates the trained flow matching-based model as a deterministic regularizer.
	This approach replaces the stochastic and highly iterative updates required by score-based or diffusion-based models with a deterministic and computationally efficient update derived from the learned velocity field, thus significantly accelerating the channel estimation process.
	\item 
	Extensive numerical simulations on the 3rd generation partnership project (3GPP) standardized clustered delay line (CDL) channel datasets \cite{3GPP} demonstrate that the proposed estimator achieves superior accuracy and robustness compared to recent score-based and diffusion-based SOTA methods.
	Our approach achieves up to 49-fold faster inference and 20-fold lower peak graphics processing unit (GPU) memory usage, while maintaining performance improvements across all considered scenarios.
	Notably, compared with the best performance achieved by SOTA baselines, our method attains comparable performance while requiring a 20\% reduction in pilot overhead.
	To ensure reproducibility and foster further research, we make our code and trained models publicly available~\cite{FMChannelCode}.
\end{itemize}

\section{System Model and Problem Formulation}
We consider a narrowband MIMO communication system which includes a transmitter equipped with $N_{\mathrm t}$ antennas and a receiver equipped with $N_{\mathrm r}$ antennas. 
The wireless channel between the transmitter and the receiver is characterized by the matrix $\mathbf{H} \in \mathbb{C}^{N_{\mathrm r} \times N_{\mathrm t}}$.
To perform channel estimation, the transmitter needs to transmit a set of $N_{\mathrm p}$ pilot symbols.\footnote{
	We follow prior studies \cite{Arvinte2023,Chen2025,Zhou2025} that assume the transmitter sends pilot symbols  and estimation is done at the receiver. 
	Under time division duplex (TDD) scheme and exploiting channel reciprocity, the channel estimated in downlink can be used to design uplink tranmissions. The presented framework also applies when receiver sends pilot symbols and the transmitter estimates the channels.}
Let $\mathbf{p}_i \in \mathbb{C}^{N_{\mathrm t}}$ denote the $i$-th pilot symbol, the received signal vector $\mathbf{y}_i \in \mathbb{C}^{N_{\mathrm r}} $ corresponding to the $i$-th pilot symbol is 
\begin{align}
	\mathbf{y}_i = \mathbf{H} \mathbf{p}_i + \mathbf{n}_i,
\end{align}
where $\mathbf{n}_i \in \mathbb{C}^{N_{\rm r}}$ is additive white Gaussian noise (AWGN) whose elements are independent and identically distributed (i.i.d.) complex Gaussian random variables with zero mean and variance $\sigma^2 $.

Let $\mathbf{P} = [\mathbf{p}_1, \mathbf{p}_2, \dots, \mathbf{p}_{N_{\rm p}}] \in \mathbb{C}^{N_{\rm t} \times N_{\rm p}}$ denote the matrix that concatenates the $N_{\rm p}$ pilot symbols.
The received signal matrix  $\mathbf{Y}\in \mathbb{C}^{N_{\rm r} \times N_{\rm p}}$, which is observed during the pilot transmission period, is given by
\begin{align}\label{eq:2}
	\mathbf{Y} = \mathbf{H} \mathbf{P} + \mathbf{N},
\end{align}
where $\mathbf{N} = [\mathbf{n}_1, \mathbf{n}_2, \dots, \mathbf{n}_{N_{\rm p}}]  \in \mathbb{C}^{N_{\rm r} \times N_{\rm p}}$ is the AWGN matrix.
The pilot density $\alpha$ is defined as $\alpha = N_{\rm p} / N_{\rm t}$.
In practice, a smaller $\alpha$ is preferred since it lowers pilot overhead during channel estimation and leaves more resources for payload data.

Given  information of the transmitted pilot symbols $\mathbf{P}$, channel estimation aims to infer the  channel matrix $\mathbf{H} $ from the noisy pilot measurement $\mathbf{Y}$.
We denote the probability density functions of  $\mathbf{H}$ and $\mathbf{Y}$ as $p_H(\cdot)$ and $p_Y(\cdot)$, respectively.
Following Bayesian principles, the estimated channel matrix \raisebox{-0.35ex}{$\hat{\mathbf{H}}$} can be obtained by solving the following maximum a posteriori (MAP) estimation problem:
\begin{align}\label{eq:map_posterior}
	\hat{\mathbf{H}} 
	& =  \underset{\mathbf{H}} {\argmax}~\log p_{H}(\mathbf{H}|\mathbf{Y}) \\\notag
	& =  \underset{\mathbf{H}}{\argmax}~\log p_{Y}(\mathbf{Y}|\mathbf{H}) +  \log p_{H}(\mathbf{H}),
\end{align}
where $\log p_{H}(\mathbf{H}|\mathbf{Y})$ is the log-posterior, $\log p_{Y}(\mathbf{Y}|\mathbf{H})$ is the log-likelihood of the noisy measurement given the channel, and $\log p_{H}(\mathbf{H})$ is the log-prior distribution that encapsulates prior knowledge about the channel.

To make the MAP problem in \eqref{eq:map_posterior} tractable, it is customary to recast it as a regularized optimization by reframing its objective function \cite{Bishop2006}.
Since the noise matrix $\mathbf{N}$ consists of i.i.d. complex Gaussian-distributed elements with zero mean and variance $\sigma^2$, maximizing the log-likelihood $\log p_Y(\mathbf{Y}\mid\mathbf{H})$ is equivalent to minimizing the data-fidelity term defined by $\frac{1}{2\sigma^2}\,\|\mathbf{Y}-\mathbf{H}\mathbf{P}\|^2$ \cite{Kay1993}.
The log-prior $\log p_H(\mathbf{H})$ is generally intractable for realistic channel models and is therefore commonly replaced by a regularizer term $\mathcal{R}(\mathbf{H})$ that encodes prior knowledge of the ground-truth channel distribution.
By substituting the data-fidelity term and the regularizer term into \eqref{eq:map_posterior}, the log-posterior maximization problem can be rewritten as the following regularized minimization problem:
\begin{align}\label{eq:re_prob}
	\hat{\mathbf{H}} = \underset{\mathbf{H}} {\argmin}~\frac{1}{2\sigma^2} \|\mathbf{Y} - \mathbf{H} \mathbf{P} \|^2 +   \mathcal{R}(\mathbf{H}).
\end{align}

The effectiveness of the converted problem in \eqref{eq:re_prob} is critically dependent on the choice of the regularization function $\mathcal{R}(\mathbf{H})$.
Classical approaches rely on explicit priors, where $\mathcal{R}(\mathbf{H})$ is a hand-crafted regularizer designed to enforce specific properties\cite{Bajwa2010,Huang2019,Taheri2011}.
These explicit regularizers impose an overly simplified structure that often fails to capture the rich complexity of real-world channel distributions.
SOTA channel-estimation methods employ diffusion-based and score-based models to learn an implicit prior from channel data \cite{Ma2024, Arvinte2023, Chen2025, Zhou2025}. 
While these approaches achieve impressive results, they can be computationally intensive and require complex hyper-parameter tuning due to reliance on posterior-sampling algorithms (e.g., Langevin dynamics) or variational inference schemes, thereby diverging from the well-structured regularized minimization paradigm.

A robust way to solve \eqref{eq:re_prob} is using first-order proximal splitting methods \cite{Combettes2011}.
The main idea of proximal splitting methods is to decompose the original problem into two subproblems: a data-fidelity subproblem and a regularization subproblem.
In the channel estimation problem, the data-fidelity subproblem ensures that the relationship between the estimated channel $\hat{\mathbf{H}}$ and the received signal matrix $\mathbf{Y}$ is specified by the signal transmission model, while the regularization subproblem refines the estimate according to the learned statistical distribution of real-world channels.
In this context, the PnP paradigm offers a transformative way to address the regularization subproblem \cite{Venkatakrishnan2013}.
The core insight of the PnP method is to replace the proximal operator  applied to $ \mathcal{R}(\mathbf{H})$ by a powerful trained generative model.
In the following section, we leverage the PnP methodology to design a novel channel estimator powered by a denoiser derived from the flow matching-based generative model.

\section{Flow Matching Based Channel Estimator}

\subsection{Flow Matching}

Flow matching is a cutting-edge generative modeling technique that learns a  transformation from a simple source distribution $\hat{q}$ to a complex target distribution $q$ \cite{Lipman2023, Albergo2022, Liu2022c}. 
A flow $f_t$ is a time-continuous bijective transformation that maps a sample $x_0$ drawn from $\hat{q}$ (i.e., $x_0 \sim \hat{q}$) to a target $x_1 \sim q$ over the continuous time interval $t\in[0,1]$, with $f_0 (x_0)= x_0 $ and $f_1 (x_0) = x_1$.
Imagine each sample from the source distribution as a particle, the flow is the path this particle takes over the continuous time interval $t\in[0,1]$, to reach a position characteristic of the target data distribution. 
The collective movement of all particles defines a continuous evolution of the probability distribution itself, from  $\rho_0 = \hat{q}$ at $t=0$ to $\rho_1=q$ at $t=1$. 
This sequence of distributions over time, $(\rho_t)_{0\leq t \leq 1}$, is referred to as the probability path.
This entire process aforementioned is governed by a time-dependent velocity field $v_t(x)$, which dictates the direction and speed of each particle at any point $(t,x)$ in time and space, along the probability path. 
Formally, a time-dependent flow $f_t$ can be characterized by a time-dependent velocity field $v_t$ by the following ordinary differential equation (ODE) 
\begin{align}\label{eq:ode}
	\frac{\mathrm{d} f_t(x)}{\mathrm{d}t} = v_t\left(f_t\left(x\right)\right), \quad v_0 (x)= x.
\end{align}

The probability path $(\rho_t)_{0\leq t \leq 1}$ can be generated by the velocity field $v_t$ if the corresponding flow $f_t$ satisfies
\begin{align}
	f_t (x_0) \sim \rho_t\quad {\rm for}\quad x_0 \sim \hat{q}.
\end{align}

By integrating the flow dynamics in \eqref{eq:ode} from  $t=0$  to $t=1$, we can transform any initial sample $x_0 \sim \hat{q}$ to a final sample $x_1 = f_1 (x_0)$ that resembles the target data distribution $q$.
The purpose of flow matching is to build a velocity field $v_t$ whose induced flow $f_t$ can generate a probability path $(\rho_t)_{0\leq t \leq 1}$ with $\rho_0 = \hat{q}$ and  $\rho_1 = q$.
Unlike diffusion models which relies on stochastic noising and a learned reverse process, flow matching directly learns a velocity field along the probability path, yielding a deterministic sampler and greater flexibility to encode domain priors.

\subsection{Training a Flow Matching Model for MIMO Channels}

Given access to a training dataset of channel matrices $\{ \mathbf{H}_i\}_{i=1}^N$ with $N$ samples, we can use a deep neural network $\psi_\theta$ parameterized by weights $\theta$ to approximate velocity field.
In this work, $\psi_\theta$ is implemented with a U-Net architecture \cite{Ronneberger2015a}. 
Since the standard U-Net operates on real valued inputs, we represent each complex matrix $\mathbf{H}_i$ by stacking its real and imaginary parts along the feature dimension to obtain a real-valued three dimensional tensor $\tilde{\mathbf H}_i \in \mathbb R^{2 \times N_{\rm r} \times N_{\rm t}}$. 
We define the stacking operator $S\!:=\!\mathbb{C}^{N_{\mathrm{r}} \times N_{\mathrm{t}}} \!\rightarrow \!\mathbb{R}^{2 \times N_{\mathrm{r}} \times N_{\mathrm{t}}} $ as
\begin{align}\label{eq:r2c}
	\tilde{\mathbf H}_i = S(\mathbf H_i)
	&\triangleq
	\begin{bmatrix}
		\Re\!\left(\mathbf H_i\right) \\[2pt]
		\Im\!\left(\mathbf H_i\right)
	\end{bmatrix},
\end{align}
where  $\Re(\cdot)$ and $\Im(\cdot)$ denote the real part and the imaginary part of a complex matrix, respectively. 

\begin{algorithm}[h]
	\caption{Training of Flow Matching Model for MIMO Channel Estimation}
	\label{alg:fm_tr}
	\SetKwInOut{Require}{Require}
	\SetKwInOut{Output}{Output}
	\Require {Training dataset $\{\tilde{\mathbf H}_i\}_{i=1}^N$, batch size $B$, neural network $\psi_\theta$, number of epochs $N_{\text{epoch}}$, number of iterations per epoch $N_{\text{step}}$ }
	\Output{Trained parameters $\theta$}
	\For{$e=1$ \KwTo $E$}{
		\For{$n=1$ \KwTo $N_{\text{step}}$}{
			Sample $\{\mathbf x_1^{(i)}\}_{i=1}^{B}$ from dataset $\{\tilde{\mathbf H}_i\}_{i=1}^N$ and draw $\{\mathbf x_0^{(i)}\}_{i=1}^{B}$ whose entries are i.i.d.  $\mathcal{N} (0,1)$\;
			Sample $t^{(i)} \sim \mathrm{Unif}[0,1]$ for $i=1,\ldots,B$\;
			Compute $\mathbf x_t^{(i)}\!\leftarrow\!(1-t_i)\mathbf x_0^{(i)}+t_i\mathbf x_1^{(i)}$\;
			Compute the training loss according to \eqref{eq:tr_loss}\;
			Update $\theta$ via gradient descent on the training loss\;
		}
	}
\end{algorithm}

A primary challenge in training the neural vector field $\psi_\theta$ lies in the unknown nature of the ideal target velocity field $v_t$. 
Consequently, a direct regression objective like $\mathbb{E}_{t,\mathbf{x}} \left[\left\|\psi_\theta(\mathbf{x},t)-v_t(\mathbf{x})\right\|_2^2\right]$ is infeasible, since the ground truth $v_t$ is not available for evaluation. 
Conditional flow matching circumvents this issue by instead training the neural network  $\psi_\theta$ to match the velocity of a simple conditional path defined between a source sample $\mathbf{x}_0$ and a target sample $\mathbf{x}_1$.
This work utilizes the straight-line path, a choice that is highly advantageous due to its mathematical simplicity \cite{Liu2022c}. 
For any time $t\in [0,1]$, the position on this path is defined by the interpolation $\mathbf{x}_t = (1-t)\mathbf{x}_0 + t\mathbf{x}_1$.
Consequently, the velocity vector  can be derived analytically by differentiating $\mathbf{x}_t$ with respect to $t$, i.e., $ \frac{d\mathbf{x}_t}{dt} = \mathbf{x}_1 - \mathbf{x}_0$.

During training, the parameters of the neural vector field $\psi_\theta$ are optimized to minimize the conditional flow matching loss between the predicted velocity and the target velocity that drives the straight-line flow between coupled pairs. 
The procedure involves drawing a mini-batch of $B$ sample pairs $\{(\mathbf x_0^{(i)},\mathbf x_1^{(i)})\}_{i=1}^{B}$, where $\mathbf x_0^{(i)} \in \mathbb R^{2 \times N_{\rm r} \times N_{\rm t}}$ has i.i.d. entries drawn from a Gaussian distribution with zero mean and unit variance $\mathcal{N}(0, 1)$, and $\mathbf x_1^{(i)}$ is drawn from the training dataset $\{\tilde{\mathbf H}_i\}_{i=1}^N$.
The conditional flow matching loss is \cite{Lipman2023}
\begin{align}\label{eq:tr_loss}
	\mathcal L(\theta)
	=\frac{1}{B}\sum_{i=1}^{B}
	\left\|\,\psi_\theta \left(\mathbf{x}_t^{(i)},t^{(i)}\right)
	-\left(\mathbf x_1^{(i)}-\mathbf x_0^{(i)}\right)\,\right\|^2,
\end{align}
in which 
\begin{align}
	\mathbf{x}_t^{(i)} = \mathbf(1-t^{(i)})\mathbf{x}_0^{(i)} + t^{(i)}\mathbf{x}_1^{(i)} , \quad t^{(i)}\sim \mathrm{Unif}[0,1],
\end{align}
where $ \mathrm{Unif}[0,1]$ denotes the uniform distribution over the interval $[0,1]$.
The pseudo-code of the complete training process is summarized in  Algorithm~\ref{alg:fm_tr}.

\subsection{MIMO Channel Estimation via Flow Matching}
Given a well-trained velocity field $\psi_\theta$ learned with the straight line path, we build a time dependent denoiser $D_t(\mathbf{x})$, which moves an intermediate channel estimate to its ground-truth channel realization along the learned velocity, as follows \cite{Hurault2022, Ryu2019}:
\begin{align}
	D_t(\mathbf{x}) = \mathbf{x} + (1-t)\,\psi_\theta(\mathbf{x},t),
\end{align}
where $\psi_\theta(\mathbf{x},t)$ approximates the velocity at position $\mathbf{x}$ that reaches $\mathbf{x}_1$ in the target distribution $\{\tilde{\mathbf H}_i\}_{i=1}^N$.
For example,  the oracle  velocity at $\mathbf{x}_t$ is given by $\psi_\theta(\mathbf{x}_t,t) = \frac{\mathbf{x}_1-\mathbf{x}_t}{1-t}$, hence $D_t(\mathbf{x}_t) = \mathbf{x}_t + (1-t)\,\psi_\theta(\mathbf{x}_t,t) =\mathbf{x}_1$.

Using the denoiser $D_t$ defined above, we adopt the  PnP-PGD approach to perform channel estimation \cite{Venkatakrishnan2013, Hurault2022, Ryu2019}.
PnP-PGD is a first-order iterative method for solving the problem in \eqref{eq:re_prob}, where each iteration consists of a gradient descent step on the data-fidelity term followed by a denoising operation.
In our flow matching-based PnP-PGD framework, the iterations are organized over $K$ discrete time steps uniformly sampled on $t\in[0,1]$ with $t^{(k)}=k/K, k = 0, \dots, K$. 
At each time $t_k$, given the current estimated channel $\mathbf{H}^{(k)}$, we first apply a gradient descent step on the data-fidelity term in \eqref{eq:re_prob} with the received pilots matrix $\mathbf Y$ and the  transmitted pilot matrix $\mathbf P$  as follows:
\begin{align}\label{eq:update_z}
	\mathbf Z^{(k+1)} = \mathbf H^{(k)} + \gamma_k  \tfrac{\left(\mathbf Y - \mathbf H^{(k)}  \mathbf P\right) \mathbf{P}^H }{\sigma^{2}},
\end{align}
where $\gamma_k$ is the step size and $\mathbf{P}^H$ denotes the Hermitian  transpose of $\mathbf{P}$.

Then, we apply flow matching model guided denoising in the real-valued three dimensional tensor and map back to the complex-valued channel matrix as follows:
\begin{align}\label{eq:update_ch}
	{\mathbf H}^{(k+1)} = S^{-1}  \left(\tilde{\mathbf H}^{(k+1)}   \right),
\end{align}
in which 
\begin{align}\label{eq:denoise}
	\tilde{\mathbf H}^{(k+1)} = D_{t^{(k)}}(\tilde{\mathbf Z}^{(k+1)}),
\end{align}
and 
\begin{align}\label{eq:lin_it}
	\tilde{\mathbf Z}^{(k+1)} =   S\left((1 - t^{(k)})\boldsymbol{\varepsilon}^{(k)} + t^{(k)} \mathbf Z^{(k+1)}\right),
\end{align}
where $S^{-1}(\cdot)$ denotes the inverse transformation defined in \eqref{eq:r2c} that maps a real valued tensor to a complex valued matrix, and \raisebox{-0.35ex}{$\boldsymbol{\varepsilon}^{(k)} \in \mathbb{C}^{N_{\rm r} \times N_{\rm t}}$} has i.i.d entries following a complex Gaussian distribution with zero mean and unit variance.

\begin{algorithm}[h]
	\caption{Channel Estimation via Flow Matching}
	\label{alg:fm_ev}
	\SetKwInOut{Input}{Inputs}
	\SetKwInOut{Output}{Output}
	\Input{Received pilots $\mathbf Y$, pilot matrix $\mathbf P$, number of iterations $K$.}
	\Output{Estimated channel matrix $\hat{\mathbf H}$.}
	\BlankLine
	\textbf{Initialize:} $\mathbf H^{(0)} \sim \mathcal{CN}(\mathbf 0, \mathbf I)$\;
	\For{$k = 0, 1, \ldots, K-1$}{
		Compute $\mathbf Z^{(k+1)}$ according to \eqref{eq:update_z} \;
		Update  $\mathbf H^{(k+1)}$ according to \eqref{eq:update_ch}-\eqref{eq:lin_it}\;
	}
	\textbf{Return:} $\hat{\mathbf H} = \mathbf H^{(K)}$\;
\end{algorithm}

The summary of our formulated flow matching-based channel estimation method is presented in Algorithm~\ref{alg:fm_ev}.
The theoretical convergence analysis of PnP-PGD method has been established in \cite{Ryu2019, Hurault2022, Martin2024} and is omitted here due to space limitations.
It is worth noting that such theoretical guarantees are missing in recent works that use generative models such as score models and diffusion models to solve the channel estimation problem \cite{Arvinte2023, Zhou2025, Chen2025, Ma2024}.
Previous generative models based channel estimation approaches based on score and diffusion models operate by adding random noise in many small stages and then learning to remove it.
The number of stages and the noise level at each stage, known as the noise schedule, are fixed during training and are usually kept at testing \cite{Arvinte2023, Zhou2025, Chen2025, Ma2024}.
From a different route, flow matching learns a time dependent velocity that points from the current estimate toward the target channel  and the step count can be adapted to the available budget.

\section{Numerical Results}

In the experiments, we focus on a point-to-point MIMO system with \(N_{\mathrm t}=64\) transmit antennas and \(N_{\mathrm r}=16\) receive antennas operating at a carrier frequency of \(40~\mathrm{GHz}\).
Both the transmitter and receiver employ uniform linear arrays with element spacing of $\lambda/2$, where $\lambda$ denotes the carrier wavelength.
The wireless propagation channels are generated following the CDL channel models defined in the 3GPP TR~38.901 specification \cite{3GPP}.  
A total of \(10{,}000\) CDL-C channel realizations are used for training, while \(100\) CDL-C and \(100\) CDL-D realizations are reserved for testing.  
In the 3GPP specification, the CDL-C model contains only non-line-of-sight components, while the CDL-D model represents a LOS channel with a dominant LOS component and additional NLOS multipath components.
This configuration allows evaluating both in-distribution (CDL-C) and out-of-distribution (CDL-D)  performance. 
The pilot matrix \(P\in\mathbb C^{N_{\mathrm t}\times N_{\mathrm p}}\) is composed of unit-power quadrature phase shift keying (QPSK) symbols that are independently and uniformly generated.
All channel realizations are normalized using the average channel power computed over the entire training set across all samples and entries.
The average SNR is defined as $\mathrm{SNR}= {N_{\mathrm t}}/{\sigma^{2}}$.  
The channel estimation performance is measured by the normalized mean square error (NMSE), which is given by
\begin{align}
	{\rm NMSE~[dB]} = 10 \log_{10} \frac{\| \hat{\mathbf{H}}-{\mathbf{H}} \|^2}{ \|{\mathbf{H}} \|^2 }.
\end{align}

We benchmark the proposed flow matching-based channel estimation approach against the following methods:
\begin{itemize}
	\item \textbf{Score-Langevin}~\cite{Arvinte2023}: This method employs a score-based generative model that learns the gradient of the log-prior of wireless channels and estimates channel through annealed Langevin dynamics.
	\item \textbf{Score-VI}~\cite{Chen2025}:  
	Building upon the score-based model training framework of~\cite{Arvinte2023}, this approach performs channel estimation via a variational inference (VI) method.
	\item \textbf{Diffusion}~\cite{Zhou2025}:  
	A diffusion model-based channel estimator trained with the standard noise-prediction loss of the denoising diffusion probabilistic model~\cite{Ho2020} and using the posterior-sampling framework as in~\cite{Meng2024}.
	\item \textbf{Approximate MMSE:}  
	The ideal minimum mean squared error (MMSE) estimator is \(\mathbb{E}[\mathbf{H}|\mathbf{Y}]\). 
	Since the true log-prior distribution of CDL channels does not admit a closed-form expression, we approximate this bound empirically by averaging multiple posterior samples obtained with Algorithm~\ref{alg:fm_ev}. 
	Following the approach in \cite{Arvinte2023}, we generate $4$ independent posterior samples per test instance to estimate an empirical upper bound in our experiments.
\end{itemize}

We restrict the baselines to score models and diffusion models, since prior studies have shown that classical and other deep generative approaches, such as generative adversarial networks (GANs) and variational autoencoders (VAEs), perform substantially worse under similar channel estimation conditions~\cite{Arvinte2023, Zhou2025, Chen2025}.
To ensure a fair comparison, our proposed method and all baseline models are  using the same U-Net backbone with three resolution levels and feature channel widths of 32, 64, and 128, resulting in a total of approximately 3.6 million parameters \cite{Ronneberger2015a}.
During training, all models are trained for $N_{\text{epoch}} = 400$ epochs, each comprising $N_{\text{step}} = 311$ iterations, with a batch size of $32$. 
The Adam optimizer is employed with a learning rate of $1 \times 10^{-4}$~\cite{Kingma2015}.
The number of iterations used during inference is set to $K=100$.
All baselines are trained under identical settings, including network architecture, batch size, number of epochs, and optimizer configuration, to ensure a fair comparison.  
In addition, we consider a compact variant of our method, \textbf{Ours-Lite}, which uses a U-Net backbone with a total of 0.19 million parameters.
Training and inference are conducted on a Windows laptop with an NVIDIA RTX 3500 Ada Laptop GPU and an Intel Core i9-14900HX CPU using PyTorch $2.3.1$ \cite{pytorch}.
All implementation code, including the baselines and the complete CDL datasets used for training and evaluation, is publicly available on GitHub~\cite{FMChannelCode}.


\begin{figure}[htbp]
	\centering
	\includegraphics[width=0.83\linewidth]{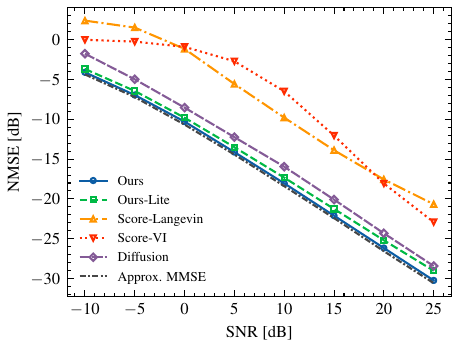}
	\caption{%
		Performance comparison on CDL-C channels.
	}
	\label{fig:snr_cdl_c}
\end{figure}

We first evaluate the in-distribution performance of our proposed channel estimator on a testing dataset generated from CDL-C channels that match the training dataset.
\figref{fig:snr_cdl_c} depicts the NMSE performance on the CDL-C channel testing dataset with pilot density $\alpha = 1$ over the  range of SNR from \num{-10} \rm{dB} to \num{25} \rm{dB}.
Our proposed method closely approaches the performance of the Approximate MMSE estimator in both low and high SNR regions.
It is clear that the flow matching-based methods, i.e., Ours and Ours-Lite, outperform other generative model-based channel estimators under all considered SNR conditions.
Specifically, our proposed method achieves at least 1.98 dB and 4.08 dB improvements compared to the diffusion model-based method and the score model-based methods when the SNR is below 0 dB, respectively.
This highlights the significant advantage of our approach in challenging low SNR scenarios.
When SNR is $25$ dB,  our proposed method maintains a performance improvement of at least $1.81$ dB over prior generative model-based channel estimation frameworks.
Despite using fewer parameters, the gap between Ours and the Ours-Lite ranges from 0.42 dB to 1.22 dB, which suggests that flow matching-based methods remain efficient without scaling the backbone network size.


\begin{figure}[htbp]
	\centering
	\includegraphics[width=0.83\linewidth]{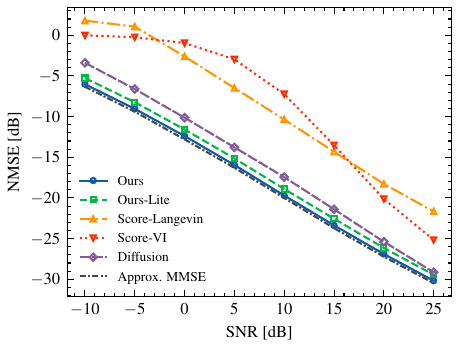}
	\caption{%
		Performance comparison on CDL-D channels.
	}
	\label{fig:snr_cdl_d}
\end{figure}

We further examine the generalization capability of the proposed channel estimator on the CDL-D channel testing dataset, which represents a different propagation configuration compared with the training dataset.
\figref{fig:snr_cdl_d} illustrates the NMSE results for pilot density $\alpha = 1$ across the SNR range from $-10~\text{dB}$ to $25~\text{dB}$.
Even under this distribution shift, our approach remains close to the Approximate MMSE bound and retains strong robustness.
For SNR from $-10$ to $0~\text{dB}$, our proposed estimator yields an average gain of $2.44$ dB over the diffusion model-based method and $8.75$ dB over the score model-based methods, respectively.
At higher SNR from $15$ to $25~\text{dB}$, it outperforms the diffusion-based and score-based estimators by $1.51$ dB and $7.21$ dB on average.
The average performance gap between Ours and Ours-Lite is $0.78$ dB, which is consistent with the CDL-C channel testing dataset case and suggests that a compact backbone network preserves the generalization capability of the flow matching estimator.


\begin{figure}[!htbp]
	\centering
	\includegraphics[width=0.83\linewidth]{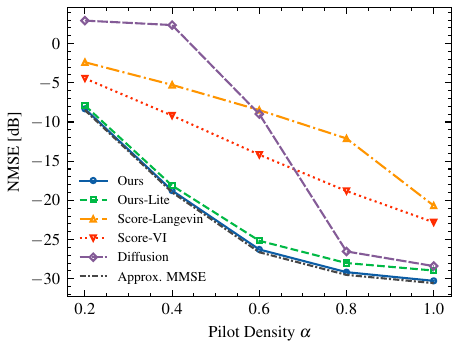}
	\caption{%
		Impacts of pilot density $\alpha$ on NMSE performance.
	}
	\label{fig:pd_cdl_c}
\end{figure}

\figref{fig:pd_cdl_c} demonstrates the NMSE performance as a function of pilot density $\alpha$ on the CDL-C dataset, at a fixed SNR of $25$ dB.
While the performance of all frameworks improves with increased pilot density, our flow matching-based methods consistently outperform the diffusion-based and score-based channel estimators.
Notably, our method demonstrates superior pilot efficiency.
For instance, it attains nearly the best NMSE achieved by the diffusion-based  and the score-based models while incurring less than 20\% and 40\% pilot overhead, respectively.
These results show that our estimator sustains superior accuracy under a  lower pilot budget, which is attractive for practical communication systems.


The runtime and peak GPU memory usage for all considered channel estimation approaches on the entire CDL-C channel testing dataset are summarized in Table~\ref{tab:runtime_memory_params}.
We monitor the peak GPU memory usage by using the PyTorch built-in function \FuncSty{torch.cuda.memory.max\_memory\_allocated} \cite{pytorch}.
Our method is approximately 5.2 times faster compared with the diffusion model based and score model based baselines.
Moreover, the Ours-Lite variant is 49 times faster and reduces peak GPU memory by 20 times while maintaining the superior accuracy.
These findings suggest that flow matching-based estimators offer a practical path for deploying generative model-based channel estimation in realistic communication systems.
\vspace{-0.3em}
\begin{table}[!ht]
	\centering
	\caption{Runtime and GPU memory usage comparison.}
	\label{tab:runtime_memory_params}
	\begin{tabular}{|l|c|c|}
		\hline
		\textbf{Method} & \textbf{Runtime [s]} & \makecell{\textbf{Peak GPU Memory [GB]}}  \\
		\hline
		{Score-Langevin}~\cite{Arvinte2023} & 438.19 & 1.75   \\
		\hline
		{Score-VI}~\cite{Chen2025}          & 127.30 & 1.75 \\ 
		\hline
		{Diffusion}~\cite{Zhou2025}         & 67.27 & 1.76 \\
		\hline
		{Approx. MMSE}                      & 63.65  & 5.76 \\
		\hline
		{Ours [Proposed]}                   & 12.89  & 1.75  \\
		\hline
		{Ours-Lite}                         & 1.37    & 0.08  \\
		\hline
	\end{tabular}
\end{table}
\vspace{-0.3em}

\section{Conclusion}
This paper presented a novel channel estimator based on the flow matching model to address the critical computational latency issue of prior diffusion-based and score-based channel estimators.
The proposed approach incorporated the velocity field learned by flow matching into the PnP-PGD framework.
Experiments on 3GPP CDL-C and CDL-D channel datasets verified that our method achieved superior accuracy, robust generalization ability to dataset from an unseen channel distribution, while simultaneously delivering up to a 49-fold reduction in inference latency and a 20-fold decrease in peak GPU memory usage. 
This work offered a practical path towards deploying generative models based channel estimation framework  in real-time communication systems.

\bibliographystyle{IEEEtran}
\bibliography{IEEEabrv,hokie-HD}
\end{document}